# Robust magnetic proximity induced anomalous Hall effect in a room temperature van der Waals ferromagnetic semiconductor based 2D heterostructure

*Hao Wu, Li Yang, Gaojie Zhang, Wen Jin, Bichen Xiao, Wenfeng Zhang, Haixin Chang\**

H. Wu, L. Yang, G. Zhang, W. Jin and B. Xiao
Center for Joining and Electronic Packaging, State Key Laboratory of Material Processing and Die & Mold Technology, School of Materials Science and Engineering, Huazhong University of Science and Technology (HUST), Wuhan 430074, China.
Institute for Quantum Science and Engineering, Huazhong University of Science and Technology (HUST), Wuhan 430074, China.

Prof. Dr. W. Zhang and Prof. Dr. H. Chang
Center for Joining and Electronic Packaging, State Key Laboratory of Material Processing and Die & Mold Technology, School of Materials Science and Engineering, Huazhong University of Science and Technology (HUST), Wuhan 430074, China.
Institute for Quantum Science and Engineering, Huazhong University of Science and Technology (HUST), Wuhan 430074, China.
Shenzhen R&D Center of Huazhong University of Science and Technology (HUST), Shenzhen 518000, China.

E-mail: hxchang@hust.edu.cn

**Abstract**

Developing novel high-temperature van der Waals ferromagnetic semiconductor materials and investigating their interface coupling effects with two-dimensional topological semimetals are pivotal for advancing next-generation spintronic and quantum devices. However, most van der Waals ferromagnetic semiconductors exhibit ferromagnetism only at low temperatures, limiting the proximity research on their interfaces with topological semimetals. Here, we report an intrinsic, van der Waals layered room-temperature ferromagnetic semiconductor crystal, $FeCr_{0.5}Ga_{1.5}Se_4$ (FCGS), with a Curie temperature as high as 370 K, setting a new record for van der Waals ferromagnetic semiconductors. The saturation magnetization at low temperature (2 K) and room temperature (300 K) reaches 8.2 emu/g and 2.7 emu/g, respectively. Furthermore, FCGS possesses a bandgap of approximately 1.2 eV, which is comparable to the widely used commercial silicon. The FCGS/graphene heterostructure exhibits an impeccably smooth and gapless interface, thereby inducing a robust magnetic proximity coupling effect between FCGS and graphene. After the proximity coupling, graphene undergoes a charge carrier transition from electrons to holes, accompanied by a transition from non-magnetic to ferromagnetic transport behavior with robust anomalous Hall effect. Notably, the anomalous Hall effect remains robust even temperatures up to 400 K.

**1 Introduction**

Proximity couplings among two-dimensional (2D) layered van der Waals materials lead to fascinating states of matter, such as quantum anomalous Hall effect, fractional quantum anomalous Hall effect, exciton condensates, chiral quantum light generation, unconventional coupled ferroelectricity and superconductivity.[1-7] Since 2D ferromagnetism were discovered in atomic-thick van der Waals layered materials,[8-11] the research on 2D ferromagnetic materials based 2D heterostructures has increased exponentially, including current-driven magnetization switching,[12] spin tunnel field-effect transistors,[13] magnetic tunnel junction (MTJ) devices,[14] asymmetric magnetic proximity interactions,[15] spin injection and detection,[16] etc. Among these materials, 2D ferromagnetic semiconductors allow simultaneous manipulation of two degrees of freedom, charge and spin, and can theoretically achieve in-situ data processing and storage. In contrast to conventional dilute magnetic semiconductor materials, 2D van der Waals ferromagnetic semiconductors emerge as the most auspicious contenders for the forthcoming spintronic and quantum devices, primarily attributed to their exceptional layered structure and facile integrability. Nonetheless, despite the notable advancements achieved in 2D van der Waals ferromagnetic metallic materials and functional devices that surpass room temperature operation,[17-20] it is worth noting that the Curie temperature of intrinsic 2D van der Waals ferromagnetic semiconductors with substantial ferromagnetic properties mostly remains confined to low-temperature regimes.

Furthermore, investigations have demonstrated that when graphene and ferromagnetic semiconductors are coupled by proximity, the hybridization between the π orbitals in graphene and the spin-polarized d orbitals in ferromagnetic semiconductors can produce exchange interactions, thereby inducing ferromagnetism in graphene.[21,22] Most of these studies focus on graphene and traditional non-van der Waals oxide ferromagnetic materials,[21,23,24] and a small part focuses on the interface between van der Waals ferromagnetic semiconductors and graphene.[15,25,26] However, due to the low Curie temperature of van der Waals ferromagnetic semiconductors, the corresponding magnetic proximity effect is constrained to temperature regimes significantly below room temperature. The van der Waals magnetic proximity effect at room temperature has not yet been realized. Utilizing the insulating properties of two-dimensional magnetic semiconductors, the magnetic proximity effect produces pure spin-polarized carriers in graphene without interference from ferromagnetic material carriers, which offers a wealth of opportunities for exploring quantum phenomena in graphene at room temperature, also serving as an invaluable platform for interface studies.

In this study, we report for the first time an intrinsic van der Waals room-temperature ferromagnetic semiconductor crystal $FeCr_{0.5}Ga_{1.5}Se_{4.0}$ (FCGS) with a 1.2 eV band gap and Curie temperature up to 370 K. By coupling it with graphene, room temperature ferromagnetism was successfully introduced into graphene, and the anomalous Hall effect in graphene remains robust at temperatures even up to 400 K.

## 2 Results and Discussions

### 2.1 Structural and phase characterizations of FCGS

The structure of FCGS is shown in Figure 1a. It exhibits a van der Waals layered structure with a theoretical interlayer spacing of 1.266 nm. This structural configuration has been substantiated through X-ray diffraction analysis.[27] High-quality single crystals of FCGS were synthesized through a chemical vapor transport (CVT) method, as detailed in the methods section. The resulting single crystals, depicted in Figure 1b, exhibit a black, metallic luster and possess a plate-like structure. These crystals can reach dimensions of up to 1 cm. The X-ray diffraction patterns (XRD) of FCGS single crystals are shown in Figure 1c. The orientation of the entire crystal is highly consistent, featuring sharp XRD peaks exclusively from the {0 0 l} crystallographic plane family, with no discernible impurity peaks detected. This observation attests to the exceptional crystalline quality of the synthesized FCGS material. The crystal orientation of the entire crystal is very consistent, featuring sharp XRD peaks exclusively from the {0 0 l} crystallographic plane family, with no discernible impurity peaks detected. This confirms the high crystalline quality of the synthesized FCGS crystals. The interlayer spacing of FCGS is calculated from the (003) peak to be 1.27 nm, which is consistent with the theoretical value.[27] The synthesized FCGS is easily exfoliated, as shown in Figure 1d, which shows stacked FCGS nanosheets exfoliated by Scotch tape onto a $SiO_2$/Si substrate, indicating the excellent and easy-to-exfoliate layered structure of FCGS. Figure 1e presents a high-resolution transmission electron microscopy (HRTEM) image of FCGS nanosheets. When observed from the c-axis direction, FCGS reveals a hexagonally arranged lattice structure. The orientation of the white lines corresponds to the (110) plane direction, with an interplanar spacing of 0.19 nm, consistent with the theoretical value[27]. Figure 1f displays the selected-area electron diffraction (SAED) of the FCGS nanosheet. Only one set of hexagonally arranged single-crystal diffraction spots is visible, providing further evidence of the exceptional crystalline quality of the synthesized FCGS single crystal. The calculated interplanar spacing for the (110) plane is determined to be 0.19 nm, which is in agreement with the HRTEM results. Figures 1g and 1h represent the dark-field image and the corresponding elemental mapping of the FCGS nanosheet, respectively. It is evident that within the entire image area, the elements Fe, Cr, Ga, and Se are uniformly distributed across the entire nanosheet, without any discernible elemental clustering or aggregation. The corresponding energy-dispersive X-ray spectroscopy (EDS) of the FCGS nanosheet in Figure 1g is shown in Figure S1, and the respective elemental content distribution is presented in Table S1. The molar ratio of Fe, Cr, Ga, and Se elements is approximately 0.99: 0.51: 1.54: 3.96, indicating that the molecular formula of FCGS is $FeCr_{0.5}Ga_{1.5}Se_4$. Figure 1i presents a high-angle annular dark-field scanning transmission electron microscopy (HAADF-STEM) cross-sectional image of a FCGS nanosheet. It offers a clear view of FCGS's layered structure. The measured interlayer spacing of FCGS, determined to be 1.27 nm, is in agreement with the XRD results. The Raman spectra of FCGS with varying thicknesses are depicted in Figure 1j. FCGS exhibits thicknesses ranging from the bulk to 7 layers (bulk, 106 L, 55 L, 49 L, 34 L, 15 L, 11 L, 7 L). Corresponding atomic force microscopy (AFM) images and thickness profiles are presented in Figure S2a-g. FCGS exhibits a characteristic peak at 200 cm$^{-1}$, and notably, this peak position remains unchanged as the thickness decreases. We attribute this characteristic peak to the phonon mode associated with the Fe-Se bonds[28]. As the thickness is reduced to 9 nm, a broadened Raman peak emerges at 252 cm$^{-1}$. This occurrence is attributed to the enhanced chemical reactivity of FCGS under the influence of the quantum size effect, resulting in the appearance of this peak due to oxidation[29]. Figures 1k, m and Figure S3a, b respectively depict the optical images (Figure 1k, Figure S3a) and the corresponding Raman mapping (Figure 1m, Figure S3b) of FCGS single-crystal nanosheets with thicknesses of 106 L and 55 L. It can be observed that the Raman intensity is uniformly distributed across the entire surface of the single-crystal nanosheets, indicating the uniformity of FCGS single-crystal nanosheets.

To investigate the chemical states of magnetic elements within FCGS, X-ray Photoelectron Spectroscopy (XPS) analysis was performed. As illustrated in Figure 2a, only Fe 2P$_{3/2}$ (711.7 eV) and Fe 2P$_{1/2}$ (725.6 eV) peaks, along with their corresponding satellite peaks, were detected. This indicates that the Fe element exists in the form of

$Fe^{3+}$ ions, and there are no magnetic metallic Fe impurities within the FCGS single crystal[30]. Figure 2b presents the XPS spectrum of Cr 2P, where two distinct peaks are observed at binding energies of 587.5 eV and 577.5 eV. These peaks correspond to Cr $2P_{1/2}$ and Cr $2P_{3/2}$, respectively, and their binding energies are consistent with those expected for $Cr^{3+}$ ions[31,32]. Figure 2c exhibits the XPS spectrum of Ga 2P, where the Ga $2P_{1/2}$ and Ga $2P_{3/2}$ peaks are located at 1145.2 eV and 1118.2 eV, respectively. These binding energies are in accordance with the expected values for Ga 2P in Ga-Se system[33]. Figure 2d represents the X-ray Photoelectron Spectroscopy (XPS) spectrum of Se, with Se $3d_{5/2}$ and Se $3d_{3/2}$ peaks located at 55.5 eV and 54.6 eV, respectively. These binding energies are consistent with those commonly observed in selenides[34,35].

**2.2 Semiconductor characteristics of FCGS**

To verify the semiconductor characteristics of FCGS, a variety of analytical techniques were employed. As shown in Figure 2a, the ultraviolet-visible-near-infrared (UV-Vis-Nir) absorption spectrum of the FCGS single crystal is presented. The linear part of the plot is extrapolated to the x-axis, resulting in a bandgap width of 1.2 eV. Figure 2b displays the photoluminescence (PL) spectrum under 808 nm excitation light, with a central wavelength of the photoluminescence emission peak at 1018 nm. The corresponding bandgap width is 1.2 eV, consistent with the results obtained from UV-Vis-Nir absorption spectrum. The bandgap of FCGS is very close to that of widely used commercial silicon semiconductors[36], indicating significant application potential for FCGS. Besides optical characterization, semiconductor properties of FCGS were examined through photodetector measurements, as demonstrated in Figure 3c. A temporal photoresponse was recorded for a 19 nm-thick FCGS nanosheet under excitation with light at a wavelength of 1050 nm and an intensity of 20 mW/cm². Under these conditions, a photoresponsivity of 0.4 A/W was observed. Figure 3d illustrates the temperature-dependent resistance (R-T) characteristics of the FCGS bulk single crystal over the temperature range of 160 K to 350 K. Notably, as the temperature decreases, the resistance undergoes an exponential increase, in conformity with the well-established R-T behavior commonly observed in semiconductors[37,38]. The corresponding Arrhenius plot is depicted in Figure 3e, where it is evident that the relationship between the resistance of FCGS and temperature perfectly conforms to the Arrhenius Equation, displaying a perfect linear correlation between ln R and 1/T. The calculated thermal activation energy (Ea) is determined to be 0.23 eV, significantly smaller than the optical bandgap of 1.2 eV. The significant difference between the Ea and the bandgap suggests the possible presence of a subband gap within FCGS[39]. Figure 3f presents the current-voltage (IV) characteristics of bulk FCGS single crystals at various temperatures. It is observed that as the temperature increases, the slope of the IV curves gradually increases, consistent with the temperature-dependent resistance behavior, further confirming the semiconductor characteristics of the FCGS single crystal.

**2.3 Room temperature ferromagnetism in FCGS**

The FCGS single crystal exhibits remarkable room-temperature ferromagnetism, as depicted in Figure 4a. The figure shows zero-field-cooled (ZFC) and field-cooled (FC) curves over the temperature range of 2-400 K under an in-plane magnetic field of 0.1 T, collectively manifesting the characteristic features of typical ferromagnetic thermomagnetic curves. The Curie temperature of FCGS is approximately 370 K, marking the highest recorded value among all known van der Waals intrinsic ferromagnetic semiconductors[8,9,40,41]. Figure 4b displays the M-H magnetization curve of the FCGS crystal at 2 K in an in-plane configuration, revealing pronounced ferromagnetic hysteresis characteristics. The magnetization intensity reaches 8.2 emu/g at a magnetic field of 10 T, approximately 0.74 μB/f.u., surpassing that of most van der Waals layered ferromagnetic semiconductors[40,41]. Figure 4c displays magnetization curves within the temperature range of 50 K to 400 K. Below the Curie temperature, the curves exhibit clear hysteresis loop characteristics, while above the Curie temperature, they display linear paramagnetic magnetization curves. At room temperature (300 K), the maximum magnetization intensity can reach 2.7 emu/g, approximately 0.24 μB/f.u.. Figure 4d depicts ZFC and FC curves over the temperature range of 2-400 K, under an out-of-plane magnetic field of 0.1 T. In comparison to the in-plane mode, it is evident that the magnetization in the out-of-plane mode is significantly lower, indicating the in-plane easy magnetization axis property of FCGS. Figure 4e represents the hysteresis loop of FCGS at 2 K in an out-of-plane configuration. In comparison to the in-plane magnetization curve shown in Figure 4b, this further confirms the in-plane easy magnetization axis property of FCGS single crystals. Figure 4f displays magnetization curves in the out-of-plane configuration within the temperature range of 50 K to 400 K. In comparison to Figure 4c, it's evident that the in-plane easy magnetization axis property of FCGS single crystals is maintained at all temperatures below the Curie temperature.

To further verify the in-plane easy magnetization characteristics of FCGS, we conducted magnetic force microscopy (MFM) tests, as shown in Figure S4. Figure S4a presents an AFM image of a 90 nm thick FCGS nanosheet, while the corresponding MFM phase image is displayed in Figure S4b. Notably, the brightness at the edges of the FCGS nanosheet is significantly higher or lower than that within the plane, indicating the presence of magnetic poles at the edges of the FCGS nanosheet. This suggests that the magnetic field lines predominantly align in-plane, thus confirming the in-plane easy magnetization axis of FCGS nanosheets. Figure S4c depicts an AFM image and the corresponding MFM phase image of a 28 nm thick FCGS nanosheet. Similarly, it can be observed that the magnetic poles of this thickness of FCGS nanosheet are located at the edge, exhibiting in-plane easy magnetization axis characteristics.

## 2.4 The van der Waals interface between FCGS and graphene

The van der Waals layered structure and room-temperature ferromagnetic semiconductor properties of FCGS provide an ideal platform for studying magnetic proximity effects. Utilizing metallic ferromagnetic materials in the study of magnetic proximity effects often involves the incorporation of spin transport signals originating from the ferromagnetic metal material itself, resulting in frequently unreliable outcomes[42,43]. The utilization of room temperature ferromagnetic semiconductors can help circumvent the interference caused by the intrinsic spin transport signal of the ferromagnetic material itself. However, current research in this domain remains confined to van der Waals ferromagnetic semiconductors with low Curie temperatures, thus constraining investigations to low-temperature regimes[25,44]. Through the combination of high-resistance semiconducting FCGS with non-magnetic van der Waals metals, it becomes possible to measure the pure spin transport properties within van der Waals metals. Establishing a good contact interface between FCGS and non-magnetic van der Waals metals is crucial for investigating room-temperature magnetic proximity effects. We chose graphene as the non-magnetic metal van der Waals material, which is a Dirac semi-metallic material with good electrical conductivity. As shown in Figure 5a, a cross-sectional high-angle annular dark-field scanning transmission electron microscopy (HAADF-STEM) image of the mutually stacked FCGS and graphene (Gr) prepared via micro-transfer technique is presented. The outermost layer of FCGS shows no signs of oxidation or amorphous structure and a well-established van der Waals contact interface is formed between FCGS and Gr, as represented by the orange dashed line. Figures 5b-g depict elemental mappings of the FCGS/Gr heterojunction shown in Figure 5a. It is evident that various elements within FCGS are positioned above the van der Waals interface, whereas carbon (C) element in graphene are localized below the van der Waals interface, clearly delineating their distinct spatial arrangement. Figure 5h represents the EDS profiles of various elements along the direction of the red dashed line shown in Figure 5a. It is observable that at the van der Waals interface (at ~7 nm), there is a rapid transition of elements between FCGS and Gr. Figure 5i presents a high-resolution HAADF STEM image of the FCGS/Gr interface, while Figure 5j shows the intensity profile along the direction of the blue dashed line in Figure 5i. The gap width between FCGS layers measures approximately 0.40 nm, while that between graphene layers is about 0.34 nm. Remarkably, the gap width between FCGS and graphene is approximately 0.32 nm, which is smaller than the former two. It proves that an excellent van der Waals contact interface can be formed between FCGS and graphene, which is conducive to the formation of a robust magnetic coupling effect between FCGS and graphene.

**2.5 Robust above-room temperature magnetic proximity effect and induced AHE**

The magnetic proximity effect between FCGS and graphene can be characterized through Hall effect measurements, as depicted in Figure 6a, which is an optical image of a 4.8 nm thick graphene (14 L) Hall device, with the corresponding AFM height profile in the inset. The corresponding Raman spectrum is shown in Figure 6b. It can be seen that the clear G-band and 2D-band peaks of graphene are located at ~1590 cm$^{-1}$ and ~2720 cm$^{-1}$ respectively[45]. No defect peaks are found, indicating the high crystallinity of graphene.

Figure 2c illustrates the Hall resistance-magnetic field ($R_{xy}$-H) curves for the graphene Hall device depicted in Figure 2a, measured within the temperature range of 100-400 K. Clearly, within the magnetic field range of -2 T to 2 T, the $R_{xy}$-H curves manifest as straight lines with a negative slope, signifying that the carrier type in pure graphene is electrons, and there is no intrinsic ferromagnetism. Figure 6d depicts an optical image of a Hall device featuring a 3.6 nm-thick graphene layer (11 L) and a 28 nm-thick FCGS layer (22 L) heterojunction, accompanied by the corresponding AFM height profile in the inset. In Figure 6e, the corresponding Raman spectrum exhibits concurrent peaks at approximately 200 cm$^{-1}$, ~1590 cm$^{-1}$, and 2720 cm$^{-1}$ for FCGS and graphene, providing evidence of a van der Waals heterostructure with vertical stacking between FCGS and graphene. Figure 2f displays the $R_{xy}$-H curves of the FCGS/graphene Hall device over the temperature range of 2-400 K. It is notable that, with the introduction of FCGS, the $R_{xy}$-H curves exhibits nearly linear positive slopes within the ±1 to ±2 T magnetic field range. This observation indicates a transition in the carrier type of graphene from electrons to holes, confirming the presence of a well-established van der Waals interface between FCGS and graphene and a strong proximity coupling effect. In addition to the change in carrier type, it can be observed that after the incorporation of FCGS, the Hall resistance curve of graphene exhibits an inverted 'S'-shaped anomalous Hall resistance profile. This signifies the successful introduction of ferromagnetism into graphene. Whether at the low temperature of 2 K or the room temperature of 300 K, the anomalous

Hall effect remains robust, indicating a robust magnetic proximity coupling effect between FCGS and graphene. Even at 400 K, exceeding Tc, the anomalous Hall effect persists. We note that the anomalous Hall effect occurring above Tc can be attributed to the strong proximity effect between FCGS and graphene, as well as the possibility that the critical temperature for AHE in FCGS may be higher than the Tc defined by ZFC/FC when current injected[46,47].

## 3 Conclusions

In conclusion, we have discovered a van der Waals, layered, room-temperature intrinsic ferromagnetic semiconductor FCGS. It exhibits a Curie temperature as high as 370 K and a bandgap of approximately 1.2 eV. The magnetization at 10 K and 300 K reaches approximately 0.74 and 0.24 μB/f.u., respectively. A perfect van der Waals close contact interface is formed between FCGS and graphene, allowing for a strong proximity effect between them. After the proximity, graphene undergoes a change of the carrier type from electrons to holes, and a change from non-magnetic to ferromagnetic. The anomalous Hall effect remains stable even at a high temperature up to 400 K. Our discovery opens new possibilities for studying quantum anomalous Hall effects at room temperature and beyond.

## 4 Experimental Section

**Sample Preparation:** The single crystal bulk FCGS was synthesized using the chemical vapor transport method (CVT). Initially, 0.04 mol of Fe powder, 0.032 mol of Cr powder, 0.048 mol of Ga bulk, along with 0.16 mol of Se powder and 0.0047 mol of $I_2$, were loaded into a quartz ampoule with a diameter of 6 cm and a length of 33 cm. Subsequently, the ampoule was vacuum-sealed at a pressure of 0.1 Pa. It was then placed inside a horizontal tube furnace with two temperature zones. The source material end and crystal growth end were heated to 1000°C and 800°C, respectively, within one hour. The system was maintained at these temperatures for one week before naturally cooling down. Abundant black plate-like crystals, identified as single crystals of FCGS, were observed in the low-temperature region of the crystal growth end. The graphite single crystal bulk was sourced from HQ-graphene company. The FCGS nanosheets used in TEM in Figure 1e were obtained by liquid-phase exfoliation of bulk FCGS in an ethanol solution, followed by deposition onto TEM copper grids. FCGS nanosheets on $SiO_2$/Si substrates was mechanically exfoliated by Scotch tapes. Cross-section HAADF-STEM samples and heterojunction samples for Hall device were prepared by pressing the nanosheets from Scotch tape onto a PDMS surface, followed by positioning them onto the corresponding electrode locations on the $SiO_2$/Si substrate under optical microscopy, and finally releasing them by applying proper heat.

**Characterizations:** The structure and phase of FCGS were characterized using powder X-ray diffraction (Empyrean, PANalytical B.V.), Raman spectroscopy (LabRAM HR800, Horiba JobinYvon, excitation wavelength of 532 nm). The microstructure and morphology were characterized using field emission transmission electron microscopy (Tecnai G2 F30), spherical aberration corrected transmission electron microscopy (JEM ARM 200F, JEOL). The bandgap of the FCGS crystals was measured using a UV-Vis-NIR Spectrophotometer (SolidSpec-3700, Shimadzu) and a photoluminescence spectrometer (FLS1000, Edinburgh). The chemical valence states were determined by X-ray Photoelectron Spectroscopy (AXIS SUPRA+). The thickness and microscale magnetism of the nanosheets were measured using atomic force microscopy and magnetic force microscopy (AFM, XE7, Park). For photocurrent measurements, a Keithley 4200-SCS instrument was employed.

**Magnetic and electrical property measurements:** Thermomagnetization curves (M-T, ZFC, and FC modes) and magnetization curves (M-H) were measured using a physical property measurement system (PPMS, Dynacool, Quantum Design) equipped with a vibrating sample magnetometer (VSM).
For the M-T test, a heating/cooling rate of 2 K min$^{-1}$ was employed. For the M-H test, a magnetic field sweeping rate of 100 Oe s$^{-1}$ was used. For electrical measurements, the BRT and ETO modules of PPMS were employed. For R-T measurements, the Resistivity module of BRT was employed with a temperature ramp rate of 2 K min$^{-1}$ and a sampling interval of 1 K. For $R_{xy}$-H measurements, a magnetic field sweep rate of 100 Oe s$^{-1}$ and a sampling interval of 500 Oe were used. All resistivity measurements were conducted in constant current mode, with each data point being averaged from 25 individual measurements. For the IV measurements, the ETO module was used, and measurements were conducted after a 10-minute stabilization period at each set temperature. For the photocurrent test, Keithley 4200-SCS was employed.

## Supporting Information

The Supporting Information is available.

## Acknowledgments

This work is financially supported by the National Key Research and Development Program of China (grant no. 2022YFE0134600), the National Natural Science Foundation of China (grant nos. 52272152, 61674063 and 62074061), the Natural Science Foundation of Hubei Province, China (grant no. 2022CFA031), the Foundation of Shenzhen Science and Technology Innovation Committee (grant nos. JCYJ20180504170444967 and JCYJ20210324142010030), and China Postdoctoral Science Foundation (grant no. 2022M711234).

## Conflict of Interest

The authors declare no conflict of interest.

## Data Availability Statement

The data that support the findings of this study are available from the cor-responding author upon reasonable request

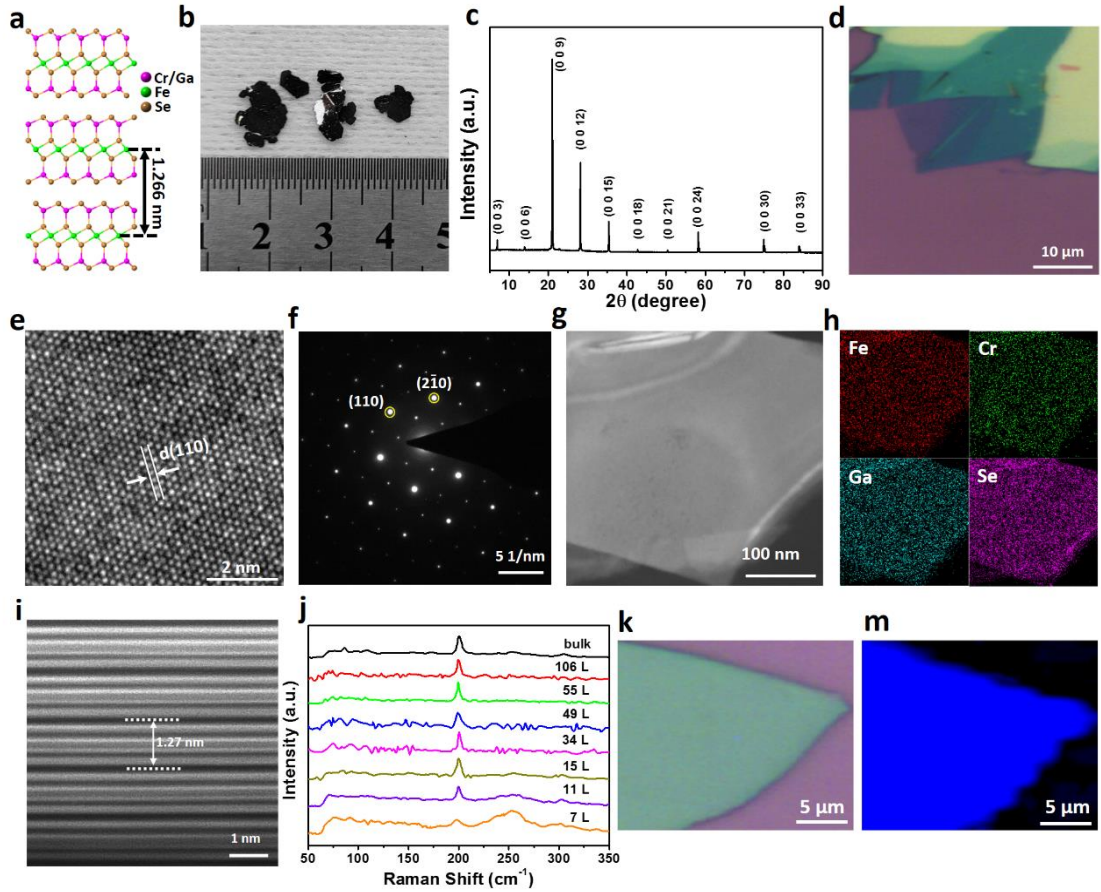

**Figure 1.** Van der Waals layered structure of FCGS. a) Crystal structure of FCGS crystal. b) Optical photographs of CVD grown FCGS crystals. c) X-ray diffraction pattern for the FCGS single crystal. d) Optical microscopy image of exfoliated FCGS nanosheets. e) High-resolution transmission electron microscopy (HRTEM) of the FCGS nanosheet. f) The selected-area electron diffraction (SAED) for the FCGS nanosheet. g-h) Dark-field microscopy and corresponding element mappings for the FCGS nanosheet. i) Cross-section HAADF STEM image of the FCGS nanosheet. j) Thickness-dependent Raman spectra for FCGS bulk and nanosheets. k-m) Optical microscopy image (k) and corresponding Raman mapping (m) for a 106 layer FCGS nanosheet.

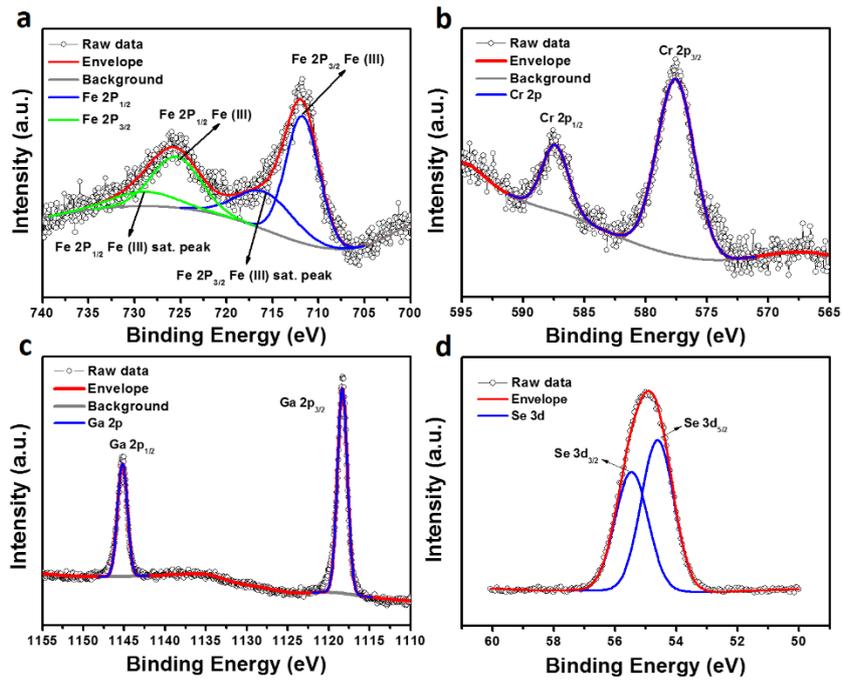

**Figure 2.** Chemical states analysis for FCGS. a-d) X-ray photoelectron spectroscopy (XPS) of Fe 2P (a), Cr 2P (b), Ga 2P (c) and Se 3d (d) for the FCGS single crystal.

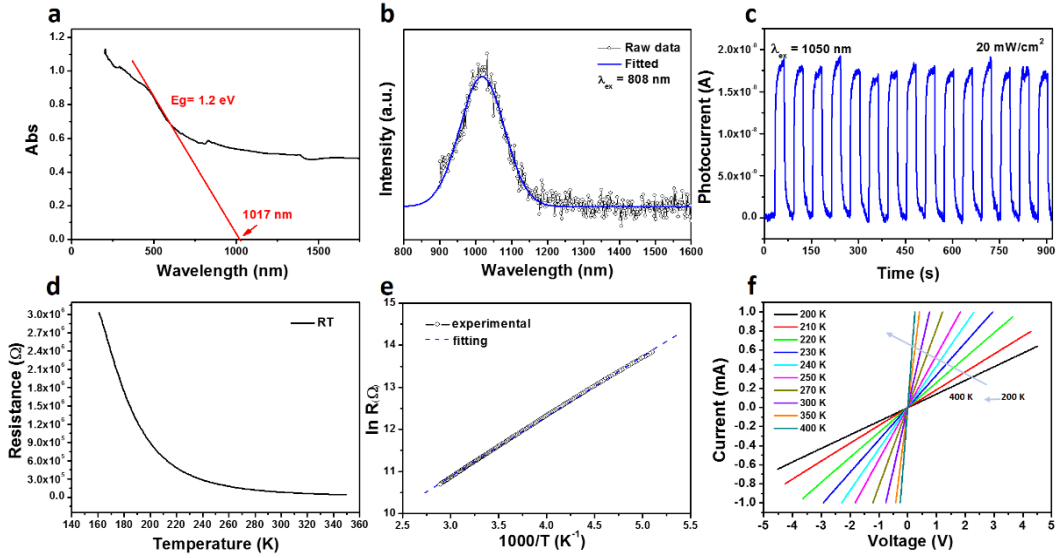

**Figure 3.** Semiconductor characteristics for FCGS crystals. a) UV-Vis-NIR absorption spectrum of bulk FCGS single crystal. b) Photoluminescence spectrum for bulk FCGS single crystal. c) Temporal photoresponse of a FCGS nanosheet. FCGS dimensions: 20 μm (length) x 12 μm (width) x 19 nm (thickness). d-f) Resistance-temperature curve (d), Arrhenius plot (e) and IV curves (f) for the bulk FCGS crystal. FCGS dimensions: 2.0 mm (length) x 1.5 m (width) x 106 μm (thickness).

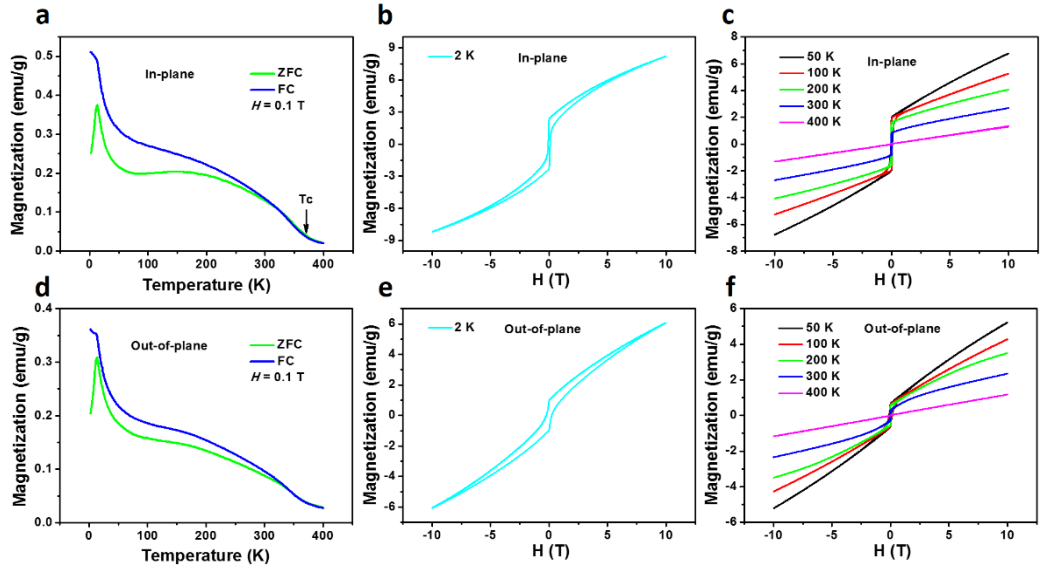

**Figure 4.** Room-temperature ferromagnetism of bulk FCGS. a) Thermomagnetic curves (ZFC and FC) with 0.1 T in-plane field applied. b) In-plane magnetization curve (M-H) at 2 K temperature. c) In-plane magnetization curves at temperatures from 50 K to 400 K. d) Thermomagnetic curves (ZFC and FC) with 0.1 T out-of-plane field applied. e) Out-of-plane magnetization curve (M-H) at 2 K temperature. f) Out-of-plane magnetization curves at temperatures from 50 K to 400 K.

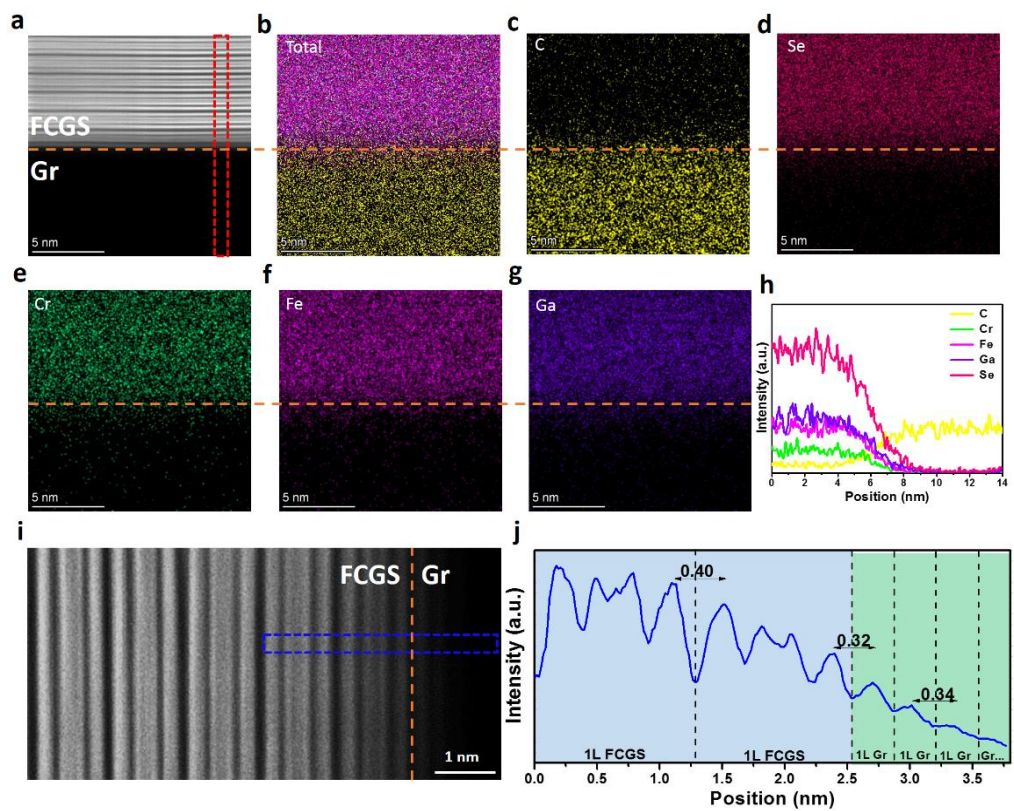

**Figure 5.** Interface between FCGS and graphene. a) Cross-section HAADF STEM image. The orange dashed line indicates the interface between FCGS and Gr. b-g) EDS element mappings of total (b), C (c), Se (d), Cr (e), Fe (f) and Ga (g). h) EDS profile along the direction of the red dashed line in (a). i) High-resolution HAADF STEM image at the interface of FCGS and graphene. j) Intensity profile along the direction of blue dashed line in (i).

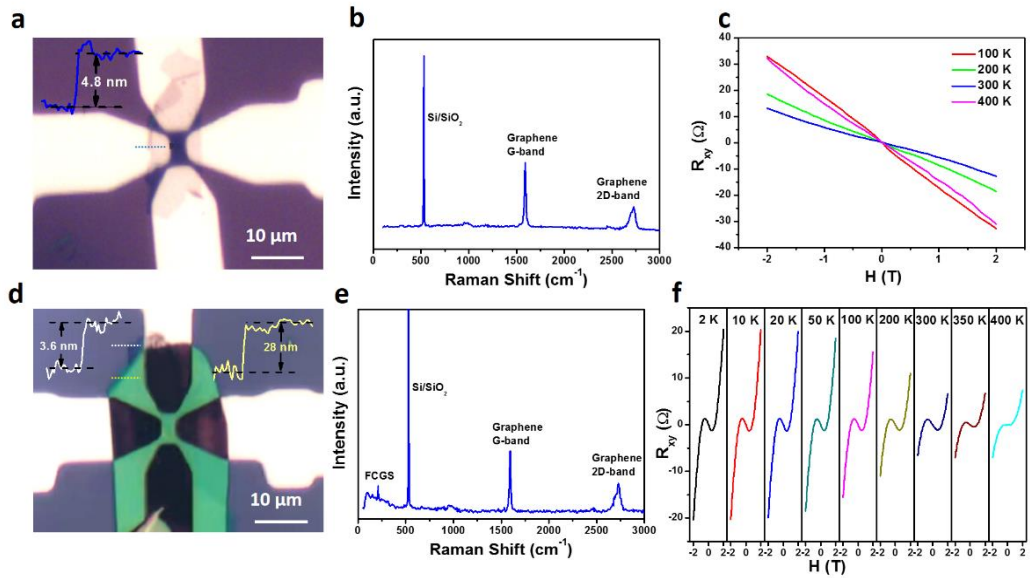

**Figure 6.** Robust magnetic proximity effect in FCGS/graphene heterostructure. a) Optical microscopy image of pure graphene Hall device. b) Raman spectrum of the graphene in (a). c) Field-dependent Hall resistance ($R_{xy}$-H) of the graphene Hall device in (a) at temperatures from 100 K to 400 K. d) Optical microscopy image of FCGS/graphene heterostructure Hall device. e) Raman spectrum of the FCGS/graphene heterostructure in (d). f) Field-dependent Hall resistance ($R_{xy}$-H) of the FCGS/graphene heterostructure Hall device in (d) at temperatures from 2 K to 400 K.